\input{aipcheck}

\documentclass[
    ,final            
  ]
  {aipproc}

\layoutstyle{6x9}

\def\etal{{\it et al.\ }}
\newcommand{\be}{\begin{equation}}  \newcommand{\ba}{\begin{eqnarray}}
\newcommand{\ee}{\end{equation}}  \newcommand{\ea}{\end{eqnarray}}

\begin{document}

\title{The Role of Turbulence in AGN Self-Regulation in Galaxy Clusters }
                                   


\author{Evan Scannapieco}{
  address={School of Earth and Space Exploration,  Arizona State University, P.O.  Box 871404, Tempe, AZ, 85287-1404, USA}
}

\author{Marcus Br\"uggen}{
  address={Jacobs University Bremen, P.O. Box 750\,561, 28725 Bremen, Germany}
}

\begin{abstract}

Cool cores of galaxy clusters are thought to be heated by low-power
active galactic nuclei (AGN), whose accretion is regulated by
feedback. However, the interaction between the hot gas ejected by the
AGN and the ambient intracluster medium is extremely difficult to
simulate, as it involves a wide range of spatial scales and gas that is
Rayleigh-Taylor (RT) unstable. Here we use a subgrid model
for RT-driven turbulence to overcome these problems and present the
first observationally-consistent hydrodynamical simulations of AGN self-regulation in galaxy
clusters.   For a wide range of parameter choices  the cluster in our
three-dimensional simulations  regulates itself for at least several $10^9$
years.  Heating balances cooling through a string of outbreaks with
a typical recurrence time of $\approx 80$ Myrs, a timescale that depends
only on the global cluster properties. 

\end{abstract}

\maketitle


\section{Introduction}

Many galaxy clusters have strong peaks in  their  X-ray surface
brightness distributions, indicating that their central gas is cooling
rapidly.  Yet spectra of such cool-core clusters show  that this
gas fails to cool below $\approx 1$ keV \cite{mcnamara:07,rafferty:06, peterson:01},  
which means that radiative cooling
must be balanced by an unknown energy source.  Currently, the most
successful model for achieving this balance relies on heating by
outflows from a central active galactic nucleus (AGN) \cite{bruggen:02,magliocchetti:07}.  Galaxies near
the centers of groups and clusters  show a boosted likelihood of
hosting AGN-driven jets \cite{best:05}. The energies  from such jets are comparable
to those  needed to balance cooling  \cite{simionescu:08}, and they
increase in proportion to the cooling  luminosity as expected in an
operational feedback loop \cite{rafferty:06, birzan:04}.

However, the full simulation of this feedback cycle remains elusive.
Direct simulation of a galaxy cluster from the Mpc scale down to the
parsec scale at which mass accretes onto the central supermassive
black hole is still out of reach even for the latest generation of
adaptive-mesh refinement (AMR) codes. Hence, in most  hydrodynamical
simulations of AGN feedback, the energy input  is set by hand, rather
than computed from the conditions surrounding the  central galaxy.  In
fact only \cite{vernaleo:06}, \cite{brighenti:06},
and \cite{cattaneo:07} attempted to simulate
self-regulated accretion, and none of these studies was able to
reproduce  the observed properties of cool-core clusters.

The biggest obstacle to these simulations is that it is unclear
how the AGN mechanical power is deposited  as thermal energy
throughout the cluster. Details of the flow depend on
the unknown viscosity of the intracluster medium (ICM), the role of
magnetic fields, and the properties of ICM turbulence -- which mixes material
on scales unresolved by current simulations.  Here we show how
tracking subgrid turbulence provides the missing piece necessary to
construct full numerical  simulations of AGN self-regulation in
cool-core clusters.  Furthermore, with no fine tuning, such
simulations naturally explain both the mechanical power of the AGN and
their duty cycles.

\section{Method}

The simulations were performed with FLASH version 3.0, a
multidimensional AMR hydrodynamics code.  While the direct simulation
of turbulence is extremely computationally-expensive and dependent on
resolution \cite{glim:01}, its behavior can be accurately approximated
with a subgrid model.  In \cite{scannapieco:08} we showed that 
AGN jets in the ICM are disrupted into resolution-dependent
pockets of underdense gas in  pure-hydro simulations, but proper modeling of subgrid
turbulence \cite{dimonte:06} shows that this a poor approximation
to a turbulent cascade that continues far beyond current resolution
limits.

The simulations presented here were also carried out using this
subgrid turbulence model, and for our overall cluster profile, we
adopted a model \cite{roediger:07} that reproduces the properties of
the Perseus cluster.   We implemented radiative cooling in the
optically-thin limit throughout the simulation, and the energy input
from AGN  was calculated from the instantaneous conditions near the
centre of the cluster.   In particular, we considered a model in which
a  fixed fraction of the gas within the central 3 kpc of the cluster
accretes onto the central supermassive black hole within a cooling
time, and a fixed fraction of the rest mass energy of this
accreted gas is returned to the ICM by increasing the temperature within
two hot spots, located $10$ kpc above and below the central AGN.
Further details of our simulations are given in \cite{bruggen:09}.

\section{Results}

\begin{figure*}

\includegraphics[width=2.5in]{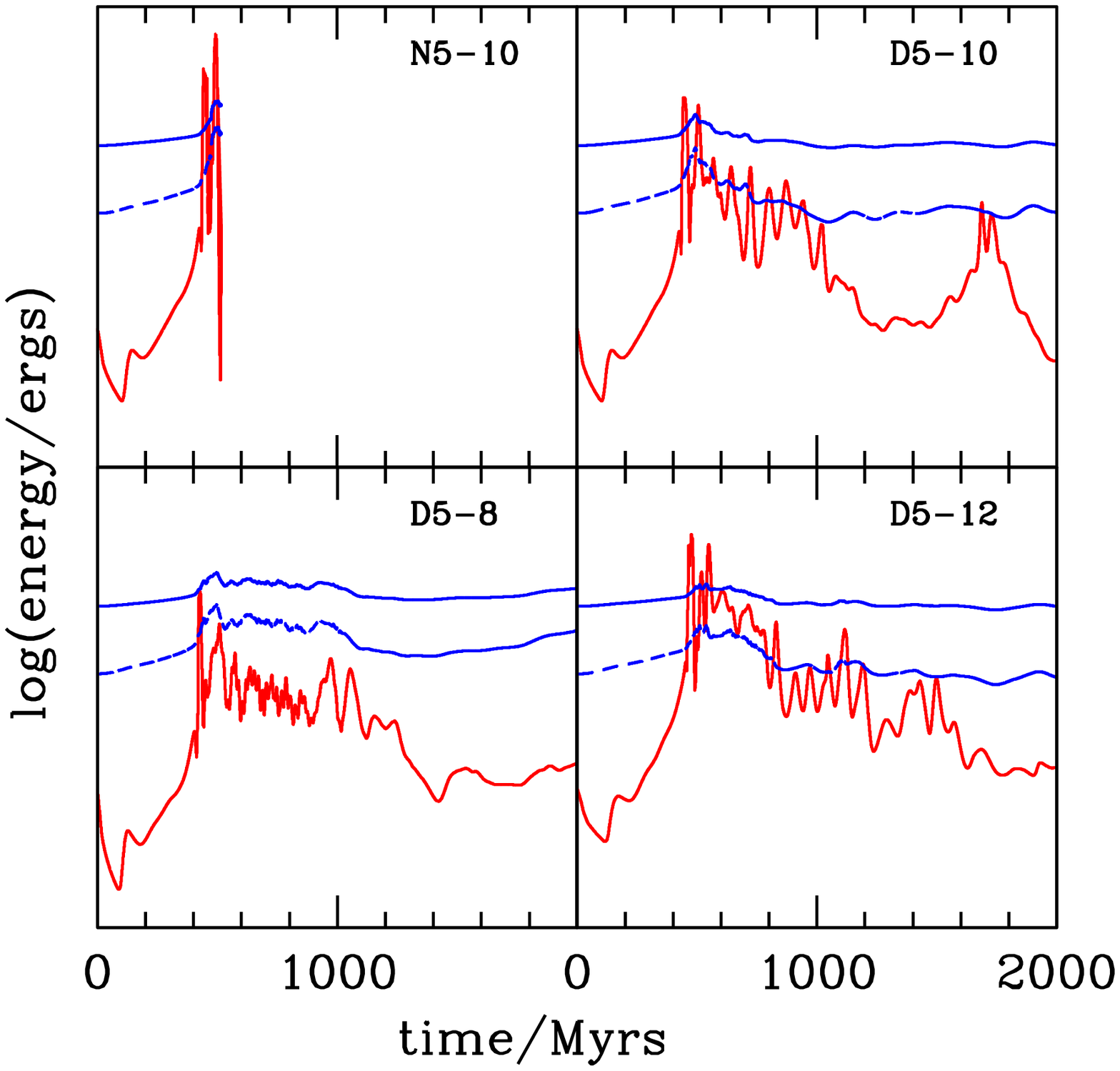}
\includegraphics[width=3.5in]{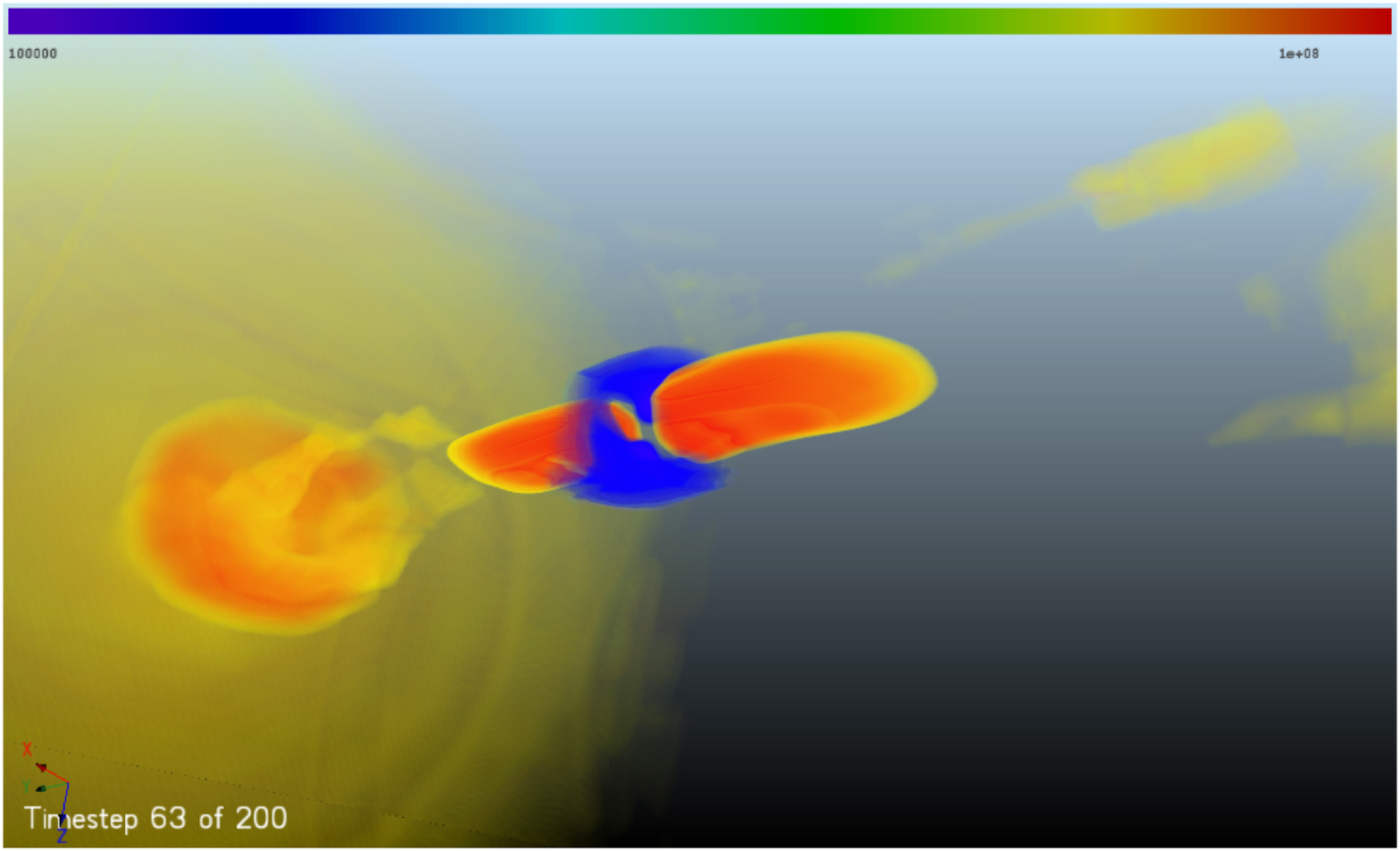}

\caption{{\em Left panels:} Jet power and integrated cooling rates for a run
without subgrid turbulence (top left), our fiducial run with subgrid
turbulence (top right), and two runs  with subgrid turbulence and
different hot-spot sizes  (bottom left and bottom right).  The solid
blue lines show the total cooling rate in the simulation volume, the
dashed blue lines show the cooling rate in the central 100 kpc,  and
the red lines show the AGN energy input.  The lines for the run
without subgrid turbulence stop abruptly at about 450 million years
because  catastrophic cooling in the center halted the
simulation. \, \,  {\em Right panel:} Volume rendering of the
temperature in our fiducial run  at $t = $ 630 Myrs. The blue ring shows
the cool gas accreting onto the center; the red and yellow blobs
represent the hot gas ejected by the AGN. On the left side one can see
an older bubble from an earlier outburst.}

\label{fig:64}
\end{figure*} 

As a control case, we first present the results of a model without
subgrid turbulence (labeled 5N-10), whose evolution is
summarized in the top left panel in Fig.\ 1.  In this case the AGN
jets do not couple to the central region.  Instead, cool gas
accumulates in the center, and  this cooling eventually leads to a
drastic AGN outburst.  However even this extreme heating does not stop the
flow of infalling gas.  Instead cooling
increases catastrophically and we are forced to stop the simulation at 480 Myrs. 

On the other hand, when the subgrid turbulence model is
switched on, as in run 5D-10, the jets couple  effectively to the
central region.  Turbulence mixes in hot material near the center
of the cluster, heating the cool inflowing intracluster gas
and stopping the AGN outburst.  Thus instead of cooling
catastrophically,  the overall radial temperature and density  profile
of the cluster remains extremely stable over the course of the
simulation, even though substantial cooling continues at all times.

Despite this overall stability,  the center of the cluster executes a
series of oscillations as indicated in Fig.\ 1.   After each outburst
of AGN activity, turbulence decays away at the eddy turn-over time scale,
 and mixing near the cluster center becomes progressively less
efficient. This leads to an increased level of accretion 
as more and more cold gas makes its way onto the AGN.  The
result is a new burst of AGN activity, which drives a jet on time scales of the
order of a sound crossing time. The RT-unstable jet leads to a rise in 
turbulent mixing, which quickly quenches
accretion when the turbulent length scale grows to be of the order of
the scale of the accretion flow.   At this point the AGN remains
relatively quiescent until the turbulence decays away again, 
repeating the cycle.  This interaction of the AGN-heated
regions with the cool inflowing gas is also illustrated in the
volume-rendered image shown in the right panel of Fig.\ \ref{fig:64}. 

It is important to point out that we did not tweak the parameters of the subgrid
model to achieve this self-regulating cycle. 
Note also that the time scale for the cycle 
is not the sound crossing time for
the central region, which is of the order of 10 Myrs.
Instead the period between the episodes of AGN
fueling is determined by the time it takes for turbulence to 
decay after it has mixed the gas in the cluster center.  
This is given by 
\be
t_{\rm duty} \approx l/v_{\rm turb},
\ee
where $l$ is the distance between the accretion region of the AGN and hot spots
 and $v_{\rm turb}$ is a typical the turbulent velocity, which we 
assume grows in a dynamical time given by $v_{\rm turb }
\approx gt \approx gl /c_s$ , where $c_s$ is the sound speed. 
Because the cluster is in hydrostatic balance, the gravitational acceleration can be written as 
$g = c_s^2\frac{1}{\rho}\frac{d\rho}{dr} \equiv c_s^2/r_0.$ This leads to the turbulent velocity  
$v_{\rm turb} = g l/c_s \approx c_s \frac{l}{r_0},$ and thus
\be  
t_{\rm duty} \approx \frac{r_0}{c_s}.
\label{eq:dutycycle2}
\ee 
This means that the size of the hot spots  does not enter the expression
for the duty cycle as would be expected in a cycle regulated by
laminar flow.   It is the properties of the cluster itself, rather
than the jet physics of the central AGN that are setting the
recurrence time of the jets.  For the  cluster simulated here, the
central sound speed is around 700 km/s and the scale height of the
cluster is around 60 kpc, so the duty  cycle is 60 kpc/700 km/s
$\approx$ 80 Myr regardless of other parameters.

To test this hypothesis we have varied the geometry of the injection
region, leading to the heating and cooling evolution shown in the
lower panels in the left of Fig.\ 1.  The time between two subsequent
outbursts for runs D5-8 and D5-12, in which the hot spots are placed at
8 and 12 kpc from the center respectively, is roughly 80 Myrs, just
as in the fiducial case.  Furthermore this timescale is consistent
with that measured by a number of observational methods 
\cite{alexander:87,owen:98,mcnamara:05,nulsen:05}.
Such cycles are also seen in a sample of nearby elliptical galaxies \cite{nulsen:09}, and it will be
interesting to compare our simulations of turbulent self-regulating AGN to
such measurements in the near future.


\begin{theacknowledgments}
MB acknowledges the support by the DFG grant BR 2026/3 within the Priority
Programme ``Witnesses of Cosmic History'' and the supercomputing grants NIC
2195 and 2256 at the John-Neumann Institut at the Forschungszentrum J\"ulich.
All simulations were conducted on the `Saguaro' cluster operated by the 
Fulton School of Engineering at Arizona State University.
The FLASH code is a product of the DOE
ASC/Alliances-funded Center for Astrophysical Thermonuclear Flashes at the
University of Chicago. 

\end{theacknowledgments}



\end{document}